\documentclass[twocolumn,showpacs,preprintnumbers,amsmath,amssymb,prb,superscriptaddress]{revtex4}
\usepackage[latin1]{inputenc}
\usepackage{dcolumn,color}
\usepackage{bm}
\usepackage{graphicx}
\usepackage[english]{babel}
\usepackage[bbgreekl]{mathbbol}
\usepackage{bbm}
\usepackage{stmaryrd}

\DeclareMathOperator{\Tr}{Tr}

\DeclareMathOperator{\diag}{diag}
\def\bbm[#1]{\mbox{\boldmath $#1$}}
\newcommand{\ket}[1]{\displaystyle{|#1\rangle}}
\newcommand{\bra}[1]{\displaystyle{\langle #1|}}
\newcommand{\TE}{\text{TE}}
\newcommand{\TM}{\text{TM}}

\makeatletter
\def\amsbb{\use@mathgroup \M@U \symAMSb}
\makeatother

\usepackage{hyperref}
\hypersetup{colorlinks,
linkcolor=black,
filecolor=green,
urlcolor=blue,
citecolor=black}

\graphicspath{{./Figures/}}

\begin{document}

\title{Radiative heat transfer between metallic gratings using adaptive spatial resolution}

\author{Riccardo Messina}
\author{Antonio Noto}
\author{Brahim Guizal}
\affiliation{Laboratoire Charles Coulomb (L2C), UMR 5221 CNRS-Universit\'{e} de Montpellier, F- 34095 Montpellier, France}
\author{Mauro Antezza}
\affiliation{Laboratoire Charles Coulomb (L2C), UMR 5221 CNRS-Universit\'{e} de Montpellier, F- 34095 Montpellier, France}
\affiliation{Institut Universitaire de France, 1 rue Descartes, F-75231 Paris Cedex 05, France}

\date{\today}

\begin{abstract}
We calculate the radiative heat transfer between two identical metallic one-dimensional lamellar gratings. To this aim we present and exploit a modification to the widely-used Fourier modal method, known as adaptive spatial resolution, based on a stretch of the coordinate associated to the periodicity of the grating. We first show that this technique dramatically improves the rate of convergence when calculating the heat flux, allowing to explore smaller separations. We then present a study of heat flux as a function of the grating height, highlighting a remarkable amplification of the exchanged energy, ascribed to the appearance of spoof-plasmon modes, whose behavior is also spectrally investigated. Differently from previous works, our method allows us to explore a range of grating heights extending over several orders of magnitude. By comparing our results to recent studies we find a consistent quantitative disagreement with some previously obtained results going up to 50\%. In some cases, this disagreement is explained in terms of an incorrect connection between the reflection operators of the two gratings.
\end{abstract}

\pacs{12.20.-m, 42.79.Dj, 42.50.Ct, 42.50.Lc}

\maketitle

\section{Introduction}

Two bodies kept at different temperatures and separated by a vacuum gap experience a radiative heat transfer mediated by photons. This energy exchange is limited by the well known Stefan-Boltzmann's law in the far field, i.e. when the distance separating the bodies is large compared to the thermal wavelength $\hbar c/k_B T$, of the order of $8\,\mu$m at ambient temperature. The pioneering works of Rytov~\cite{Rytov}, Polder and van Hove~\cite{PoldervH} first showed that this limit can be surpassed in the near-field regime, as a result of the tunneling of evanescent waves. In particular, the heat transfer can exceed the one between two blackbodies (i.e. the ideal far-field scenario predicted by Stefan-Boltzmann's law) even of several orders of magnitue when the materials support surface resonances, such as plasmons in metals (typically lying in the ultraviolet range of frequencies) and phonon-polaritons in dielectrics (typically in the infrared) \cite{JoulainSurfSciRep05}. Since the contribution of each field mode to radiative heat transfer is weighted by the Planck thermal distribution, negligible in the ultraviolet range at ordinary temperatures, dielectrics supporting surface resonances are typically the best candidates to maximize the heat flux.

Stimulated by the theoretical developments, several applications have been proposed for radiative heat transfer, ranging from thermophotovoltaic~\cite{DiMatteoApplPhysLett01,NarayanaswamyApplPhysLett03,LarocheJApplPhys06,BasuIntJEnergyRes07,FrancoeurApplPhysLett08,BasuJApplPhys09} or solar thermal~\cite{SwnasonScience09,Chen11} energy conversion, to heat-assisted data storage \cite{SrituravanichNanoLett04}, nanoscale cooling~\cite{cooling}, and the recent emerging field of thermotronics \cite{devices}. On the experimental side, the theoretical predictions have been verified during the last decade both in the plane-plane and sphere-plane configuration, for a wide range of distances, going from some nanometers to several microns~\cite{KittelPRL05,HuApplPhysLett08,NarayanaswamyPRB08,RousseauNaturePhoton09,ShenNanoLetters09,KralikRevSciInstrum11,OttensPRL11,vanZwolPRL12a,vanZwolPRL12b,KralikPRL12,KimNature15,SongNatureNano15,StGelaisNatureNano16,WatjenAPL16,KloppstecharXiv}.

Recently, the idea of manipulating the heat flux through the manipulation of some external parameters have attracted a remarkable attention. In fact, both the control of the overall value of the flux and its spectral properties can be extremely relevant for several applications, and in particular for energy conversion. In the spirit of a manipulation thourgh geometrical properties, structured surfaces have been the topic of many theoretical investigations. More specifically, by considering both 1D and 2D periodic gratings, both the radiative heat transfer~\cite{BiehsAPL11,LussangePRB12,GueroutPRB12,LiuAPL14,DaiPRB2015,DaiPRB2016a,DaiPRB2016b,YangPRL16} and the Casimir force at and out of thermal equilibrium~\cite{LambrechtPRL08,ChiuPRB10,DavidsPRA10,IntravaiaPRA12,LussangePRA12,GueroutPRA13,GrahamPRA14,WagnerPRA14,NotoPRA14,MessinaPRA15} have been studied by employing a variety of theoretical approaches. It has been shown that gratings represent indeed a tool to modify, both by reducing and amplifying, radiative heat transfer, as well as to influence its spectral properties. Concerning the Casimir force, it must be mentioned that the force between a sphere and a dielectric~\cite{ChanPRL08,BaoPRL10} or a metallic~\cite{BanishevPRL13,IntravaiaNatComm14} grating has been recently measured and theoretically investigated~\cite{MessinaPRA15}.

In the domain of radiative heat transfer, metallic gratings deserve ideed a special attention. In fact, although, as mentioned above, surface resonances for a metal typically have ultraviolet frequencies, the presence of a periodically structured pattern can result in the presence of new surface resonances, referred to a spoof plasmons~\cite{PendryScience04,GarciaVidalJOptA05}, whose frequency can be adjusted as a function of the grating parameters and can be brought, for realistic values of the grating lenghtscales, in the infrared region, thus contributing to the flux. Based on this behavior, the authors of Ref.~\onlinecite{GueroutPRB12} have recently theoretically predicted a high enhancement of the flux between two identical gold gratings. A similar study has been performed for 1D and 2D metallic gratings in Refs.~\onlinecite{DaiPRB2015,DaiPRB2016a,DaiPRB2016b}, and more recently in Ref.~\onlinecite{YangPRL16}. In these papers the reflection upon each grating has been described by means of the Fourier Modal Method (FMM)~\cite{Kim12}, equipped with the factorization rule introduced in Ref.~\onlinecite{GranetJOptSocAmA96}. It is worth stressing that, since the radiative heat transfer is calculated as an integral with respect to both frequency and wavevector, the choice of the method used to derive the grating reflection operators can drastically affect the computational time as well as the results.

This paper is precisely devoted to the study of radiative heat transfer between two gold gratings. We present and discuss a modification to the FMM technique, known as Adaptive Spatial Resolution~\cite{Granet99} (we will in the following refer to this modified method as ASR), a technique specifically introduced to accelerate the convergence. This method has also been shown to overcome the known instabilities appearing for metallic gratings~\cite{GuizalOpt}. With respect to previous works, we extend here this technique to deal with arbitrary conical incidence. Based on this approach, we provide a detailed study of the influence of the grating depth on the overall value of the flux as well as on its spectral properties. We also compare our results to part of the works mentioned above on heat transfer between metallic gratings. We show that (1) the ASR technique produces a dramatic increase of the convergence rate (thus implying a drastic reduction of computational time) and (2) the numerical results for heat transfer considerably differ from results previously obtained using the standard FMM.

The paper is structured as follows. In Sec. \ref{SecIntro} we present our physical system, describing also our notation and main definitions. In Sec. \ref{SecMethods} we discuss in detail the adaptive spatial resolution. Then, in Sec.~\ref{SecNumerics} we present our numerical results, by studying the radiative heat transfer between two gold gratings as a function of the grating height. We finally give in Sec. \ref{SecConcl} some conclusive remarks.

\section{Physical system}\label{SecIntro}

\begin{center}\begin{figure}[htb]
\includegraphics[width=8.5cm]{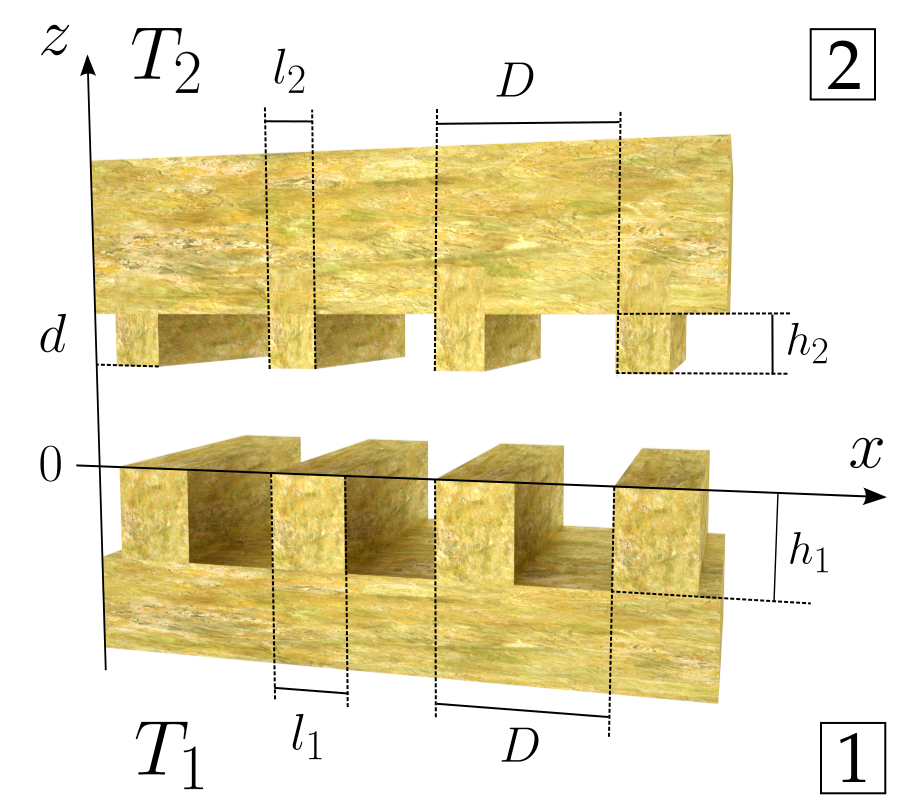}
\caption{(Color online) Geometry of the system: two gratings, labeled with 1 and 2, at a distance $d$. The gratings, in general made of different materials, are infinite in the $xy$ plane, and periodic in the $x$ direction with the same period $D$. They have corrugation depths $h_i$ ($i=1,2$), infinite thicknesses below the grating region, and lengths of the elevated part of the grating $l_i$. This defines the filling factors $f_i=l_i/D$.}\label{FigGeometry}\end{figure}\end{center}

The system we are going to address consists of two 1D lamellar gratings separated by vacuum, as shown in Fig.~\ref{FigGeometry}. The two gratings are structured along the $x$ axis, translationally invariant along the $y$ axis, and separated by a vacuum gap of thickness $d$ along the $z$ axis. The two gratings share the same period $D$, while they can have in general different filling fractions $f=l/D$ (see Fig.~\ref{FigGeometry}). Finally, they have different grating heights $h$, while the substrate below each structured region is assumed to be infinitely thick for both gratings.

The two gratings, labeled with indexes 1 and 2, are kept by some external heat source at constant temperatures $T_1$ and $T_2$. Among the numerous theoretical techniques developed to calculate the radiative heat transfer between them, we employ an approach based on the knowledge of the individual scattering operators of the bodies involved, recently introduced for systems involving two~\cite{MessinaEurophysLett11,MessinaPRA11} and three~\cite{MessinaPRL12,MessinaPRA14} bodies. This method is based on a plane-wave mode decomposition of the fields, each mode $(\omega,\mathbf{k},p,\phi)$ being identified by the direction of propagation $\phi=+,-$ along the $z$ axis, the polarization index $p$ [assuming the values $p=1,2$ which respectively correspond to transverse electric (TE) and transverse magnetic (TM) modes], the frequency $\omega$ and the transverse wavevector $\mathbf{k}=(k_x,k_y)$. In this description, the $z$ component of the wavevector $k_z$ is a dependent variable defined as
\begin{equation}\label{eq:kz}
k_z=\sqrt{\frac{\omega^2}{c^2}-\mathbf{k}^2}.
\end{equation}
In virtue of this mode decomposition, we define the trace of an operator $\mathcal{O}$ as
\begin{equation}
\Tr\mathcal{O}=\sum_p\int\frac{d^2\mathbf{k}}{(2\pi)^2}\int_0^{+\infty}\frac{d\omega}{2\pi}\bra{p,\mathbf{k}}\mathcal{O}\ket{p,\mathbf{k}}.
\end{equation}
Assuming that the external environment is thermalized with body 1, the energy this body receives per unit surface and time is given by~\cite{MessinaPRA11}
\begin{equation}
 \varphi=\hbar\Tr\Bigl[\omega n_{21}U^{(2,1)}f_{-1}(\mathcal{R}^{(2)-})U^{(2,1)\dag}f_1(\mathcal{R}^{(1)+})\Bigr],
\label{DefPhi}\end{equation}
where $\mathcal{R}^{(1)+}$ ($\mathcal{R}^{(2)-}$) is the reflection operator of body 1 (2) for an incoming wave propagating in direction $\phi=-$ ($\phi=+$) and we have defined
\begin{equation}
f_\alpha(\mathcal{R})=
\begin{cases}
\mathcal{P}_{-1}^{\text{(pw)}}-\mathcal{R}\mathcal{P}_{-1}^{\text{(pw)}}\mathcal{R}^\dag+\mathcal{R}\mathcal{P}_{-1}^{\text{(ew)}}-\mathcal{P}_{-1}^{\text{(ew)}}\mathcal{R}^\dag\\
 \hspace{5.3cm}\alpha=-1,\\
\mathcal{P}_1^{\text{(pw)}}-\mathcal{R}^\dag\mathcal{P}_1^{\text{(pw)}}\mathcal{R}+\mathcal{R}^\dag\mathcal{P}_1^{\text{(ew)}} -\mathcal{P}_1^{\text{(ew)}}\mathcal{R}\\
\hspace{5.3cm}\alpha=1.
\end{cases}
\end{equation}
Moreover, in Eq.~\eqref{DefPhi} we have defined the population differences $n_{ij}=n(\omega,T_i)-n(\omega,T_j)$, with
\begin{equation}n(\omega,T)=\frac{1}{e^{\frac{\hbar\omega}{k_\text{B}T}}-1},\end{equation}
the intracavity operator
\begin{equation}
U^{(21)}=\sum_{n=0}^{+\infty}\bigl(\mathcal{R}^{(2)-}\mathcal{R}^{(1)+}\bigr)^n=(1-\mathcal{R}^{(2)-}\mathcal{R}^{(1)+})^{-1},
\end{equation}
and the projection operators
\begin{equation}
\bra{p,\mathbf{k}}\mathcal{P}_n^\text{(pw/ew)}\ket{p',\mathbf{k}'}=k_z^n\bra{p,\mathbf{k}}\Pi^\text{(pw/ew)}\ket{p',\mathbf{k}'},
\end{equation}
where $\delta_{\phi\phi'}$ is the Kronecker delta and being $\Pi^\text{(pw)}$ [$\Pi^\text{(ew)}$] the projector on the propagative ($k<\omega/c$) [evanescent ($k>\omega/c$)] sector. We remark that, as discussed in Ref.~\onlinecite{NotoPRA14}, the periodicity on the $x$ axis makes it natural to replace the mode variable $k_x$ with
\begin{equation}k_{x,n}=k_x+\frac{2\pi}{D}n,\end{equation}
with $k_x$ taking values in the first Brillouin zone $[-\pi/D,\pi/D]$ and $n$ assuming all integer values. Based on Eq.~\eqref{DefPhi}, the calculation of the flux $\varphi$ is now reduced to the problem of describing the field reflection upon each grating. The details about the ASR technique employed in this paper are discussed in detail in the next Section.

\section{Methods: Fourier modal method with adaptive spatial resolution}\label{SecMethods}

\begin{center}\begin{figure}[b!]
\includegraphics[width=8.5cm]{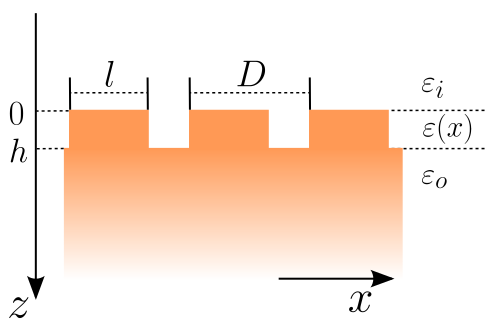}
\caption{(Color online) Geometry of the grating. Three regions are identified: $z\leq0$ and $z\geq h$ having uniform permittivity ($\varepsilon_i$ and $\varepsilon_o$, respectively) and $0<z<h$, having a periodic $\varepsilon(x)$.}\label{FigGrating}\end{figure}\end{center}

The physical system we consider is shown in Fig.~ \ref{FigGrating}: it is a lamellar grating of period $D$, invariant along the $y$ axis, with relative dielectric permittivity $\varepsilon(x)$. The grating is inlayed between two homogeneous media whose permittivities are $\varepsilon_i$ (input medium) and $\varepsilon_o$ (output medium). A monochromatic plane wave with frequency $\omega$ and parallel wavevector $\mathbf{k}=(k_x,k_y)$ illuminates the structure. Throughout the calculation, the vacuum wave number is denoted by $k_0 = \omega/c$ and we assume a time dependence of the form $e^{-i\omega t}$. The particularity of the ASR \cite{Granet99} is the use of a new coordinates system $x=F(u)$ in which the coordinate in the $x$ direction is stretched around the permittivity discontinuities. This allows to better describe  the permittivity jump, which is crucial in the case of metallic structures. Below, we derive the metric tensor and write the Maxwell's equations associated to the coordinates change. Then we solve them in the three regions of space: $z \le 0$, $0 < z < h$ and $z \ge h$. After, we derive the boundary conditions and obtain the fields amplitudes. Based on the geometry depicted in Fig.~\ref{FigGrating}, the reflection operator we are going to calculate is $\mathcal{R}^-$ for a vacuum-grating inferface located at $z=0$. This operator will allow us to easily deduce $\mathcal{R}^{(1)+}$ and $\mathcal{R}^{(2)-}$ as described in detail below.

\subsection{Solution of Maxwell's equations\\in the three regions}

We start by observing that, in general, the presence of a lamellar grating naturally divides the period region $[0,D]$ by a set of discontinuity points $\{x_l\}$ with $l=0,\dots,N$, with $x_0=0$ and $x_N=D$. In view of the change of coordinates from $x$ to $u$, we analogously define a set of points $\{u_l\}$ given by $u_l=Dl/N$, i.e. uniformly distributed between $u_0=0$ and $u_N=D$. For the explicit definition of the coordinate transformation $\left\{x=F(u), y, z\right \}$ we follow Ref.~\onlinecite{Vallius} and write
\begin{equation}\label{Transform}
F(u)=a_{1l}+a_{2l}u+\frac{a_{3l}}{2\pi}\sin \left\{ 2\pi \dfrac{u-u_{l-1}}{u_{l}-u_{l-1}} \right \},
\end{equation}
in each interval $u\in [u_{l-1},u_l]$, with $l=1,\dots,N$, where
\begin{equation}\begin{split}
a_{1l}&=(u_{l}x_{l-1}-u_{l-1}x_{l})/(u_{l}-u_{l-1}),\\
a_{2l}&=(x_{l}-x_{l-1})/(u_{l}-u_{l-1}),\\
a_{3l}&=G(u_{l}-u_{l-1})-(x_{l}-x_{l-1}).
\end{split}\end{equation}
In these expressions $G$ being a control parameter, for which we take the value $10^{-3}$ through all the numerical calculations presented in this paper.

The corresponding metric tensor is diagonal and reads $g_{ij}=\diag\left( f^2, 1, 1 \right)$, with $f(u)=dx/du=dF(u)/du$. Thus Maxwell's equations $\zeta^{ijk} \partial_jE_k=i\omega \mu_0 \sqrt{g}g^{ij}H_j$ and $\zeta^{ijk}\partial_jH_k=-i\omega \varepsilon_0 \varepsilon \sqrt{g}g^{ij}E_j$ become:
\begin{equation}\label{Eq_rot1}
\begin{cases}
\partial_y E_z-\partial_z E_y=ik_0\dfrac{1}{f(u)}\widetilde{H}_u \\ \partial_z E_u-\partial_u E_z=ik_0f(u) \widetilde{H}_y \\ \partial_u E_y-\partial_y E_u=ik_0f(u) \widetilde{H}_z \end{cases}\end{equation}
\begin{equation} \label{Eq_rot2}
\begin{cases}
\partial_y \widetilde{H}_z-\partial_z \widetilde{H}_y=-ik_0 \varepsilon(u)\dfrac{1}{f(u)}E_u \\ \partial_z \widetilde{H}_u-\partial_u \widetilde{H}_z=-i k_0 \varepsilon(u) f(u) E_y \\ \partial_u \widetilde{H}_y-\partial_y \widetilde{H}_u=-ik_0 \varepsilon(u) f(u) E_z \end{cases}\end{equation}
with $\widetilde{H}_i=Z_0H_i$, $Z_0$ being the impedance of vacuum. We now derive from Eqs.~\eqref{Eq_rot1} and \eqref{Eq_rot2} $E_z$ and $\tilde{H}_z$ and plug them into the other equations, obtained the following system for the tangential components of the fields:
\begin{equation}\label{EtoH}\begin{split}
&\partial_z \begin{pmatrix}E_u \\ E_y \end{pmatrix} \\
&=\dfrac{i}{k_0}\begin{pmatrix} -\partial_u\dfrac{1}{a(u)}\partial_y & k_0^2 f(u)+\partial_u\dfrac{1}{a(u)}\partial_u \\ -\dfrac{k_0^2}{f(u)}-\partial_y\dfrac{1}{a(u)}\partial_y & \partial_y\dfrac{1}{a(u)}\partial_u \end{pmatrix}
\begin{pmatrix} \widetilde{H}_u \\ \widetilde{H}_y \end{pmatrix},\end{split}\end{equation}
\begin{equation} \label{HtoE}\begin{split}
&\partial_z \begin{pmatrix}\widetilde{H}_u \\ \widetilde{H}_y \end{pmatrix}\\
&=
\dfrac{i}{k_0}\begin{pmatrix} \partial_u\dfrac{1}{f(u)}\partial_y & -k_0^2 a(u)-\partial_u\dfrac{1}{f(u)}\partial_u \\ \dfrac{k_0^2}{b(u)}+\partial_y\dfrac{1}{f(u)}\partial_y & -\partial_y\dfrac{1}{f(u)}\partial_u \end{pmatrix} 
\begin{pmatrix} E_u \\ E_y \end{pmatrix}, \end{split}\end{equation}
with $a(u)=f(u)\varepsilon(u)$ and $b(u)=f(u)/\varepsilon(u)$. Next, following the procedure discussed more in detail e.g. in Ref.~\onlinecite{NotoPRA14}, we rewrite this systems of equations in the Fourier space and truncate the series up to the truncation order $N$, obtaining
\begin{equation} \label{EH_Fourier}
\partial_z \begin{pmatrix}\mathcal{E}_u \\ \mathcal{E}_y \end{pmatrix}=\amsbb{F}\begin{pmatrix} \mathcal{\widetilde{H}}_u \\ \mathcal{\widetilde{H}}_y \end{pmatrix},
\qquad\partial_z \begin{pmatrix}\mathcal{\widetilde{H}}_u \\ \mathcal{\widetilde{H}}_y \end{pmatrix} = \amsbb{G} \begin{pmatrix} \mathcal{E}_u \\ \mathcal{E}_y \end{pmatrix},\end{equation}
\begin{equation} \label{FG}\begin{split}
\amsbb{F}&=
\dfrac{i}{k_0}\begin{pmatrix} k_y\alpha \llbracket a \rrbracket^{-1} & k_0^2 \llbracket f \rrbracket-\alpha \llbracket a \rrbracket^{-1} \alpha  \\ -k_0^2 \llbracket f \rrbracket^{-1}+k_y^2\llbracket a \rrbracket^{-1} & -k_y\llbracket a \rrbracket^{-1} \alpha  \end{pmatrix},\\
\amsbb{G}&=\dfrac{i}{k_0}\begin{pmatrix} -k_y\alpha  \llbracket f \rrbracket^{-1} & -k_0^2 \llbracket a \rrbracket+\alpha \llbracket f \rrbracket^{-1} \alpha  \\ k_0^2 \llbracket b \rrbracket^{-1}-k_y^2\llbracket f \rrbracket^{-1} & k_y\llbracket f \rrbracket^{-1} \alpha  \end{pmatrix},\end{split}\end{equation}
where $\alpha =\diag(k_x+2\pi n/D), n\in [-N,N]$, $k_x$ being defined in the first Brillouin zone $[-\pi/D,\pi/D]$. In Eq.~\eqref{EH_Fourier} we have gathered in the column vectors $\mathcal{E}_{u/y} $ and $\mathcal{\widetilde{H}}_{u/y}$ the $2N+1$ Fourier components of the fields $E_{u/y}$ and $\widetilde{H}_{u/y}$. Moreover, we have introduced the Toeplitz matrices $\llbracket f \rrbracket$ (resp. $\llbracket a \rrbracket$ and $\llbracket b \rrbracket$) defined in terms of the Fourier coefficients of the function $f(u)$ [resp. $a(u)$ and $b(u)$]. We can further simiplify the notation by introducing the following definitions
\begin{equation}\mathcal{E}=\begin{pmatrix}\mathcal{E}_u \\ \mathcal{E}_y \end{pmatrix}, \qquad
\mathcal{\widetilde{H}}=\begin{pmatrix}\mathcal{\widetilde{H}}_u \\ \mathcal{\widetilde{H}}_y \end{pmatrix},\end{equation}
which allow us to write
\begin{equation} \label{EqDiff}
\dfrac{\partial^2 \mathcal{E}(z)}{\partial z^2} = \amsbb{F}\amsbb{G}\mathcal{E}(z)=\amsbb{P}\amsbb{D}^2\amsbb{P}^{-1}\mathcal{E}(z).\end{equation}
In this equation, $\amsbb{D}^2$ is a diagonal matrix containing the eigenvalues of $\amsbb{F}\amsbb{G}$ and $\amsbb{P}$ is the matrix of associated eigenvectors. Then the solution of Eq.~\eqref{EqDiff} can be expressed under the form
\begin{equation} \label{SolE} \mathcal{E}(z)=\amsbb{P} \left( e^{\amsbb{D}z}\amsbb{P}^{-1}a + e^{-\amsbb{D}z}\amsbb{P}^{-1}b \right),\end{equation}
and consequently [from Eq.~\eqref{EH_Fourier}]:
\begin{equation} \label{SolH}\begin{split}\mathcal{\widetilde{H}}(z)&=\amsbb{G}\amsbb{P}\amsbb{D}^{-1} \left( e^{\amsbb{D}z}\amsbb{P}^{-1}a - e^{-\amsbb{D}z}\amsbb{P}^{-1}b \right)\\
&=\amsbb{P}'\left( e^{\amsbb{D}z}\amsbb{P}^{-1}a - e^{-\amsbb{D}z}\amsbb{P}^{-1}b \right),\end{split}\end{equation}
where $a$ and $b$ are the amplitudes of the different waves traveling in the $\phi=+$ and $\phi=-$ directions, respectively.

The calculation presented here for the grating region, characterized by a periodic $\varepsilon(x)$, can be directly applied to the homogeneous regions $z\leq 0$ and $z\geq h$ as well. In this case, the framework can be simplified by observing that $\llbracket a \rrbracket=\varepsilon\llbracket f \rrbracket$ and $\llbracket b \rrbracket=\llbracket f \rrbracket/\varepsilon$, where for the input (output) region $\varepsilon=\varepsilon_i$ ($\varepsilon=\varepsilon_o$). This directly leads to the introduction of the four additional matrices $\amsbb{P}_i$, $\amsbb{P}' _i$, $\amsbb{P}_o$, and $\amsbb{P}' _o$. In order to simplify the boundary conditions, the solution in the output medium is modified as follows
\begin{equation}\begin{split}\mathcal{E}_o(z)&=\amsbb{P}_o \left( e^{\amsbb{D}_o(z-h)}\amsbb{P}^{-1}_oa_o + e^{-\amsbb{D}_o(z-h)}\amsbb{P}^{-1}_ob_o \right),\\
\mathcal{\widetilde{H}}_o(z)&=\amsbb{P}'_o\left( e^{\amsbb{D}_o(z-h)}\amsbb{P}^{-1}_oa_o - e^{-\amsbb{D}_o(z-h)}\amsbb{P}^{-1}_ob_o \right),\end{split}\end{equation}
by means of the introduction of a phase factor.

\subsection{Boundary conditions}

Having solved Maxwell's equations in each region, we can now write down the boundary conditions at the interfaces $z=0$ and $z=h$:
\begin{equation} \label{Boundary_0} z=0 \Rightarrow
 \begin{cases}
a_{i} + b_{i}=a + b\\ \amsbb{P}'_{i} \amsbb{P}_{i}^{-1}\left( a_{i} - b_{i} \right)=\amsbb{P}'\amsbb{P}^ {-1}\left( a - b \right)
\end{cases}\end{equation}
and, denoting $\Phi=e^{h\amsbb{D}}$,
\begin{equation} \label{Boundary_h} z=h \Rightarrow
 \begin{cases}
\amsbb{P} \left( \Phi \amsbb{P}^ {-1}a + \Phi^{-1}\amsbb{P}^ {-1}b \right)=a_{o} + b_{o}\\
\amsbb{P}' \left(\Phi\amsbb{P}^ {-1} a - \Phi^{-1}\amsbb{P}^ {-1}b \right)=\amsbb{P}'_o\amsbb{P}_{o}^{-1} \left( a_o - b_o \right)
\end{cases}\end{equation}
In terms of the $\amsbb{S}$-matrix algorithm, this gives
\begin{equation} \label{Smatrix_1}\begin{split}
\begin{pmatrix} b_i \\ \amsbb{P}^ {-1}a \end{pmatrix}&=
\begin{pmatrix} -\mathbb{1} & \amsbb{P} \\ \amsbb{P}'_i\amsbb{P}_i^{-1}  & \amsbb{P}'\end{pmatrix}^{-1}\\
&\,\times\begin{pmatrix} \mathbb{1} & -\amsbb{P} \\ \amsbb{P}'_i\amsbb{P}_i^{-1}  & \amsbb{P}'\end{pmatrix}
\begin{pmatrix} a_i \\ \amsbb{P}^ {-1}b \end{pmatrix}=\amsbb{S}_1 \begin{pmatrix} a_i \\ \amsbb{P}^ {-1}b \end{pmatrix},
\end{split}\end{equation}
and
\begin{equation} \label{Smatrix_2}\begin{split}
\begin{pmatrix} \amsbb{P}^ {-1}b \\ a_o \end{pmatrix}&=
\begin{pmatrix} \Phi & \mathbb{0} \\ \mathbb{0} & \mathbb{1} \end{pmatrix}
\begin{pmatrix} -\amsbb{P} & \mathbb{1} \\ \amsbb{P}' & \amsbb{P}'_o\amsbb{P}^{-1}_o \end{pmatrix}^{-1}\begin{pmatrix} \amsbb{P} & -\mathbb{1} \\ \amsbb{P}' & \amsbb{P}'_o\amsbb{P}^{-1}_o \end{pmatrix}\\
&\,\times\begin{pmatrix} \Phi & \mathbb{0} \\ \mathbb{0} & \mathbb{1} \end{pmatrix}
\begin{pmatrix} \amsbb{P}^ {-1}a \\ b_o \end{pmatrix}=\amsbb{S}_2 \begin{pmatrix} \amsbb{P}^ {-1}a \\ b_o \end{pmatrix}.
\end{split}\end{equation}
In the last two equations the unknowns are the amplitudes $a_i$ and $b_o$ of the incoming waves, $b_i$ and $a_o$ of the reflected and transmitted waves, and the amplitudes $a$ and $b$ of the field in the grating region, in which we have absorbed the factor $\amsbb{P}^{-1}$. This is irrelevant for our purposes, since we are only interested in the field amplitudes in the homogeneous media. The final part of the calculation is straightforward (see e.g. Ref.~\onlinecite{NotoPRA14}). We define a chained $\amsbb{S}$ matrix as~\cite{granetguizalfff}
\begin{align}
\amsbb{S}=\amsbb{S}_1\circledast\amsbb{S}_2,
\end{align}
having introduced the associative operation $\amsbb{A}=\amsbb{B}\circledast\amsbb{C}$, which for three square matrices $\amsbb{A}$, $\amsbb{B}$ and $\amsbb{C}$ of dimension $4(2N+1)$ is defined as
\begin{equation}\begin{split}
\amsbb{A}_{11}&=\amsbb{B}_{11}+\amsbb{B}_{12}(\mathbb{1}-\amsbb{C}_{11}\amsbb{B}_{22})^{-1}\amsbb{C}_{11}\amsbb{B}_{21},\\
\amsbb{A}_{12}&=\amsbb{B}_{12}(\mathbb{1}-\amsbb{C}_{11}\amsbb{B}_{22})^{-1}\amsbb{C}_{12},\\
\amsbb{A}_{21}&=\amsbb{C}_{21}(\mathbb{1}-\amsbb{B}_{22}\amsbb{C}_{11})^{-1}\amsbb{B}_{21},\\
\amsbb{A}_{22}&=\amsbb{C}_{22}+\amsbb{C}_{21}(\mathbb{1}-\amsbb{B}_{22}\amsbb{C}_{11})^{-1}\amsbb{B}_{22}\amsbb{C}_{12},
\end{split}\end{equation}
where each matrix have been decomposed in four square blocks of dimension $2(2N+1)$. This $\amsbb{S}$ matrix satisfies the relation
\begin{equation}
\begin{pmatrix} b_i \\ a_o \end{pmatrix}=\amsbb{S} \begin{pmatrix} a_i \\ b_o \end{pmatrix},
\end{equation}
thus its upper-left $2(2N+1)$ block, relating the reflected field $b_i$ to the incident one $a_i$ in the upper region, can be identified as the $\mathcal{R}^-_u$ operator we are looking for in the $(u,y)$ reference system.

\begin{center}\begin{figure}[t!]
\includegraphics[width=8cm]{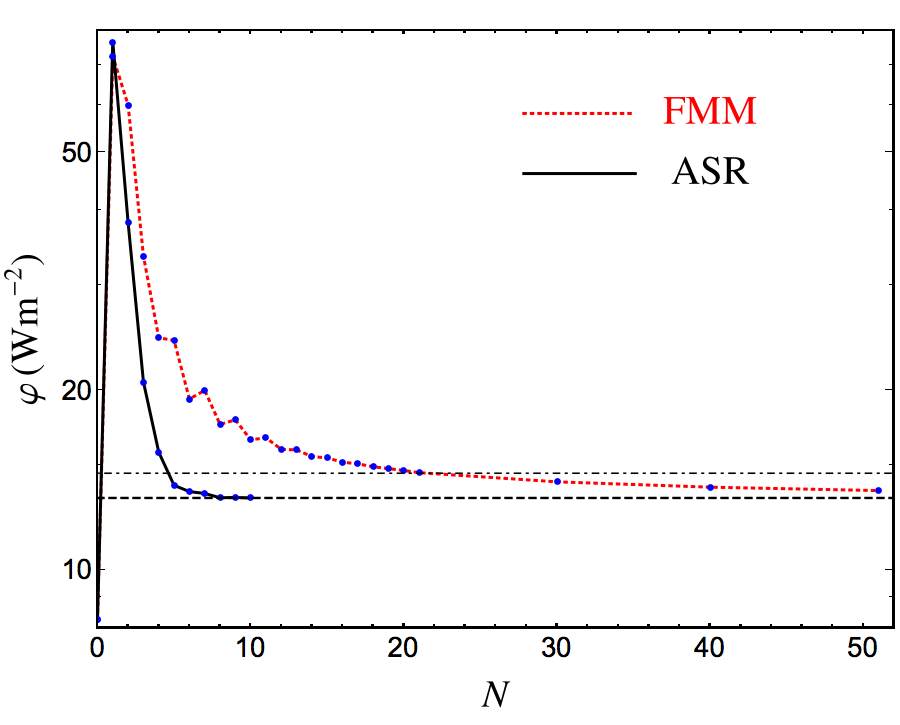}
\caption{(Color online) Heat flux between two identical gratings of height $h=2\,\mu$m, period $D=1\,\mu$m, filling fraction $f=0.5$, infinite thickness below the grating region, and temperatures $T_1=310\,$K and $T_2=290\,$K. The flux is calculated using both the FMM (red dashed line) and the ASR (black solid line), for different truncation orders $N$. The black dashed line corresponds to the asymptotic value obtained using the ASR for $N=10$, while the gray dot-dashed line is associated to a 10\% error with respect to this asymptotic value.}\label{FigFMMASR}\end{figure}\end{center}

\subsection{Transformation matrices}

The steps described in the last Section lead to the derivation of a reflection matrix $\mathcal{R}^-_u$, expressed in the transformed reference system $(u,y)$. Actually, the operator $\mathcal{R}^-$ needed for the calculation of the flux given by Eq.~\eqref{DefPhi} has to be expressed not only in the standard $(x,y)$ Cartesian reference, but also with respect to the basis of two polarization unity vectors
\begin{equation}\begin{split}
\hat{\bbm[\epsilon]}_\TE^{\phi}(\mathbf{k}_n,\omega)&=\frac{1}{k_n}(-k_y\hat{\mathbf{x}}+k_{x,n}\hat{\mathbf{y}}),\\
\hat{\bbm[\epsilon]}_\TM^{\phi}(\mathbf{k}_n,\omega)&=\frac{c}{\omega}(-k_n\hat{\mathbf{z}}+\phi k_{z,n}\hat{\mathbf{k}}_n),
\end{split}\label{PolVect}\end{equation}
where $\hat{\mathbf{a}}=\mathbf{a}/|\mathbf{a}|$ and we have defined $\mathbf{k}_n=(k_{x,n},k_y)$ and $k_{z,n}=\sqrt{\omega^2/c^2-\mathbf{k}_n^2}$. Thus, the final reflection matrix $\mathcal{R}$ takes the form
\begin{equation}
 \mathcal{R}^-=(\amsbb{B}^-)^{-1}\amsbb{T}\mathcal{R}^-_u\amsbb{T}^{-1}\amsbb{B}^+,
\end{equation}
where $\amsbb{T}$ is the matrix associated with the transformation from $(u,y)$ to $(x,y)$, whereas $\amsbb{B}^\phi$ accounts for the transition from the (TE,TM) basis to the canonical $(x,y)$ basis for fields propagating in the $\phi$ direction (note that we have $\phi=+$ for the incident field and $\phi=-$ for the reflected one, as manifest from Fig.~\ref{FigGrating}).

In order to derive $\amsbb{T}$, we start observing that we have, labeling throughout this section with a prime the fields in the $(u,y)$ coordinate system, $E'_u=f(u)E_x$ and $E'_y=E_y$. This means that the $\amsbb{T}$ matrix will be block-diagonal and take the form
\begin{equation}\amsbb{T}=\begin{pmatrix} \amsbb{T}_x & \mathbb{0} \\ \mathbb{0} & \amsbb{T}_y.\end{pmatrix}\end{equation}
Moreover, since the coordinate stretch leaves the period unchanged, one just needs to express the $2N+1$ vectors $\mathcal{E}_x$ and $\mathcal{E}_y$ as a function of $\mathcal{E}'_u$ and $\mathcal{E}'_y$ for a given value of $\omega$ and $k_x$ in the first Brillouin zone, as $\mathcal{E}_x=\amsbb{T}_x\mathcal{E}'_u$ and $\mathcal{E}_y=\amsbb{T}_y\mathcal{E}'_y$. Starting from the Fourier decompositions
\begin{equation}
E_x=\sum_ne^{ik_{x,n}x}E_{x,n},\qquad E'_u=\sum_me^{ik_{x,m}u}E'_{u,m},
\end{equation}
we easily get
\begin{equation}
 [\amsbb{T}_x]_{n,m}=\frac{1}{D}\int_0^Ddu\,e^{i[k_{x,m}u-k_{x,n}x(u)]},
\end{equation}
and in an analogous way
\begin{equation}
 [\amsbb{T}_y]_{n,m}=\frac{1}{D}\int_0^Ddu\,f(u)e^{i[k_{x,m}u-k_{x,n}x(u)]}.
\end{equation}

Concerning the second transformation matrix $\amsbb{B}^\phi$, it represents a basic change of basis and its action is written as
\begin{equation}
\begin{pmatrix} \mathcal{E}_x \\ \mathcal{E}_y \end{pmatrix}=\amsbb{B}^\phi \begin{pmatrix} \mathcal{E}^\phi_\text{TE} \\ \mathcal{E}^\phi_\text{TM} \end{pmatrix}.
\end{equation}
where the (TE,TM) basis (and thus the matrix $\amsbb{B}^\phi$) depends on the propagation direction $\phi$. Using Eq.~\eqref{PolVect}, the transformation matrix reads
\begin{equation}\amsbb{B}^\phi=
\begin{pmatrix} -\diag\Bigl(\frac{k_y}{k_n}\Bigr) & \diag\Bigl(\frac{c\phi k_{x,n}k_{z,n}}{\omega k_n}\Bigr) \\ \diag\Bigl(\frac{k_{x,n}}{k_n}\Bigr) & \diag\Bigl(\frac{c\phi k_yk_{z,n}}{\omega k_n}\Bigr) \end{pmatrix}.
\end{equation}

\subsection{Reflection operators of the two gratings}

As stated at the beginning of our calculation, the $\mathcal{R}^-$ operator we calculated is relative to a grating having its interface with vacuum at $z=0$. As a consequence, in order to deduce the $\mathcal{R}^{(2)-}$ relative to grating 2, we have to include a phase shift taking into account the fact that its vacuum-grating inferface coincides with the plane $z=d$. As described e.g. in Ref.~\onlinecite{NotoPRA14}, the matrix elements of this modified operator are given by
\begin{equation}\begin{split}
&\bra{p,\mathbf{k},n,\omega}\mathcal{R}^{(2)-}\ket{p',\mathbf{k}',n',\omega'}\\
&=\exp[i(k_{z,n}+k'_{z,n'})d]\bra{p,\mathbf{k},n,\omega}\mathcal{R}^-\ket{p',\mathbf{k}',n',\omega'}.\nonumber
\end{split}\end{equation}

We now focus on the issue of deriving $\mathcal{R}^{(1)+}$ from the known operator $\mathcal{R}^-$. For simplicity, in the following we will always consider two identical gratings, i.e. having the same height $h$ and filling factor $f$. In this case, when evaluating the flux given by Eq.~\eqref{DefPhi}, it is convenient to exploit the symmetry of the configuration and calculate (for each frequency and wavevector) only once the grating reflection matrix. Nevertheless, one must keep in mind that in our formalism the unit polarization vector associated to TE polarization defined in Eq.~\eqref{PolVect} is independent of the propagation direction $\phi$ along the $z$ axis, while the $x$ and $y$ components of the TM polarization vector are proportional to $\phi$. Since the $x$ and $y$ components are the ones which are conserved in the scattering process, this implies that the reflection matrix $\mathcal{R}^{(1)+}$ coincides with $\mathcal{R}^-$ in the diagonal blocks (TE,TE) and (TM,TM), while the non-diagonal blocks (TE,TM) and (TM,TE) undergo an overall sign change. This sign-change issue is irrelevant, of course, in the case of a planar slab, since this system does not mix the two polarizations, but it must indeed be taken into account when dealing with bodies producing a coupling between TE and TM modes.

\section{Numerical results}\label{SecNumerics}

\begin{center}\begin{figure}[b!]
\includegraphics[width=8.5cm]{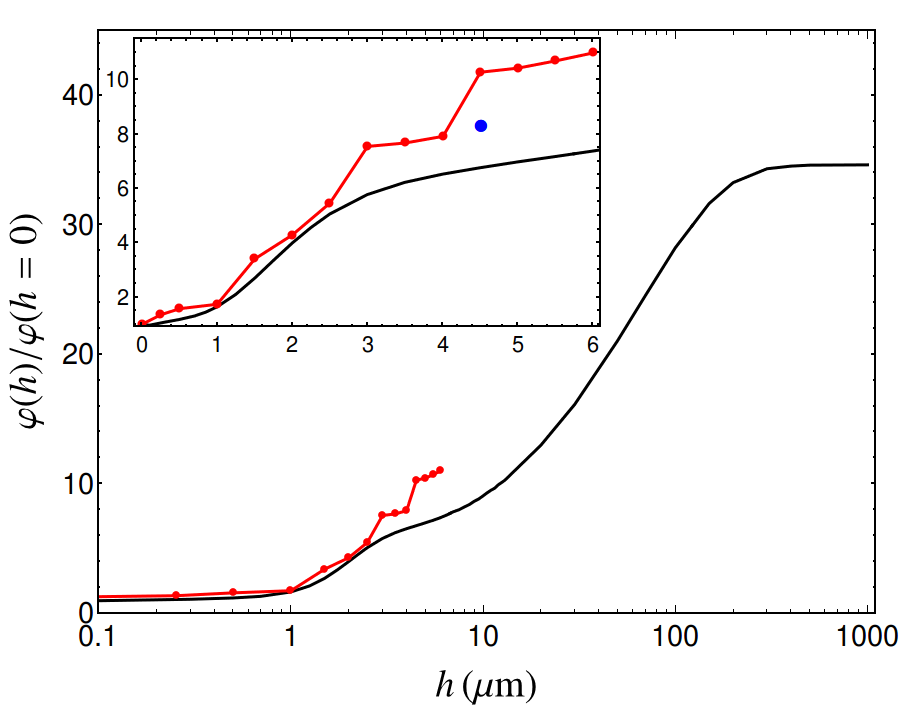}
\caption{(Color online) Ratio between the flux for a grating of height $h$ and the one for $h=0$, i.e. the one between two planar slabs. The two identical gratings have period $D=1\,\mu$m, filling fraction $f=0.5$, infinite thickness below the grating region, and temperatures $T_1=290\,$K and $T_2=310\,$K. Our results (black solid line) are compared to the results of Ref.~\onlinecite{GueroutPRB12} (red points). The blue dot corresponds to a calculation using the ASR but without inverting the sign of non-diagonal blocks of the reflection matrix (see text for more details). The inset shows a zoom of the same curve in linear $h$ scale.}\label{Figh}\end{figure}\end{center}

\begin{figure*}[t!]
\begin{center}
\includegraphics[width=18.3cm]{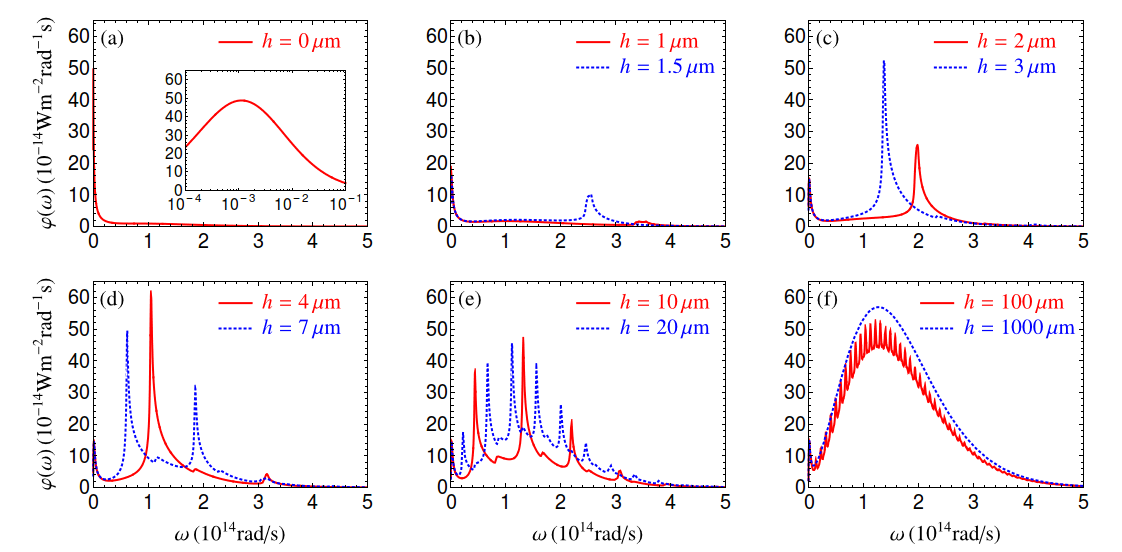}
\caption{(Color online) Spectral heat flux between two identical gratings of period $D=1\,\mu$m, filling fraction $f=0.5$, infinite thickness below the grating region, and temperatures $T_1=290\,$K and $T_2=310\,$K. In the different panels, several grating heights $h$ are represented, showing the appearance and behavior of spoof plasmons contributing to the heat flux. In panel (a), the inset show a zoom in logarithmic $\omega$ scale for small frequencies.}\label{FigGrid}
\end{center}
\end{figure*}

We are now ready to discuss our first numerical results. These concern a couple of gratings having the same features used in Ref.~\onlinecite{GueroutPRB12}. The two gratings have filling factor $f=0.5$, period $D=1\,\mu$m, temperatures $T_1=290\,$K and $T_2=310\,$K, and are placed at a distance $d=1\,\mu$m. Both gratings are made of gold, for which we have used a Drude model
\begin{equation}\varepsilon(\omega)=1-\frac{\omega_P^2}{\omega(\omega+i\gamma)},\end{equation}
where the plasma frequency and the dissipation rate are respectively equal to $\omega_P=9\,$eV and $\gamma=35\,$meV.

In order to highlight the features of the ASR technique, we start by calculating the flux for a given grating height $h=2\,\mu$m. As discussed in detail in Ref.~\onlinecite{NotoPRA14}, a crucial point when using the FMM technique is the choice of the truncation order $N$, i.e. the number of diffraction orders taken into account in the Fourier decomposition of the field, going from $-N$ to $N$. The same issue applies of course to the ASR method as well. We show in Fig.~\ref{FigFMMASR} the value of the flux $\varphi$ as a function of the truncation order, both using the FMM (red dashed line) and the ASR (black solid line). It is manifest that in both cases the first truncation orders give very different results, suggesting that $N$ has to be further increased. Nevertheless, starting from $N$ of the order of 5, the two curves are dramatically different. The curve corresponding to the ASR technique very quickly converges to a stable result, and the points corresponding to $N=8,9,10$ are basically undistinguishable. The dashed black line in the plot corresponds to the value obtained with the ASR for $N=10$, while the dot-dashed line is associated to a 10\% error with respect to this result. The results obtained with the FMM show a very different behavior. First of all, up to $N\simeq15$, the obtained flux is non-monotonic, and has a quasi-oscillatory behavior. Moreovoer, even when becoming monotonic as a function of $N$, the flux converges very slowly to its asymptotic value. More specifically, for $N=20$ the error is of the order of 10\%, while for $N=51$ (the value used in Ref.~\onlinecite{GueroutPRB12}) it is of the order of 3\%.

\begin{center}\begin{figure}[htb]
\includegraphics[width=8.5cm]{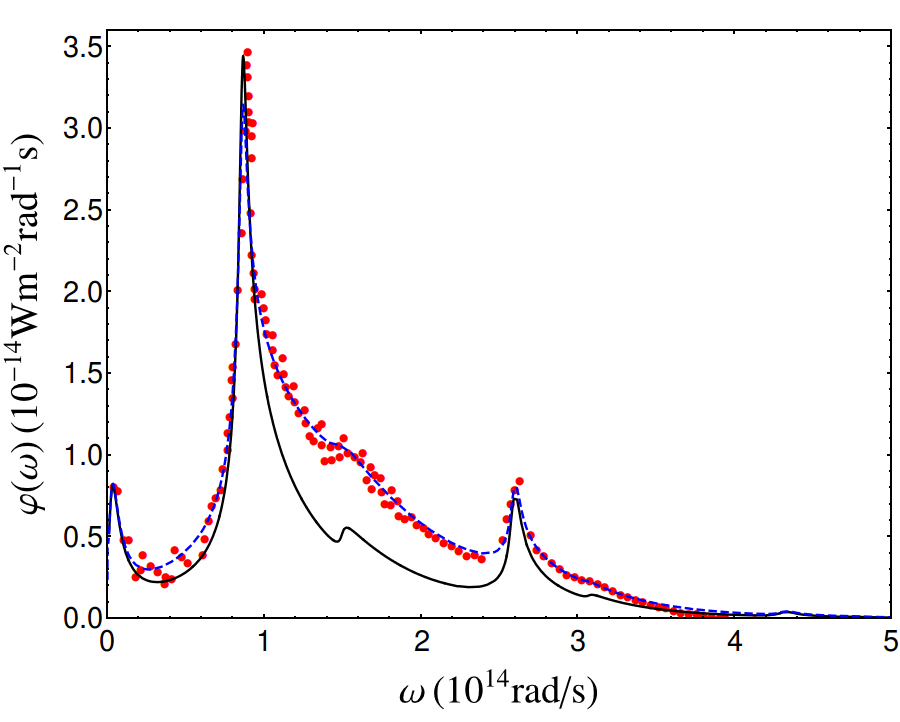}
\caption{(Color online) Spectral heat flux between two identical gratings of height $h=4.7\,\mu$m, period $D=0.5\,\mu$m, filling fraction $f=0.6$, infinite thickness below the grating region, and temperatures $T_1=300\,$K and $T_2=301\,$K. The black solid line corresponds to our calculation using the ASR, while the blue dashed line gives the result obtained without inverting the sign of non-diagonal blocks of the reflection matrix (see text for more details). The red points are extracted from Ref.~\onlinecite{DaiPhD}.}\label{FigSpectrum}\end{figure}\end{center}

Based on this discussion of the convergence, we use, in the following, a truncation order $N=8$ for the ASR, with an associated relative error of the order of 1\% on the integrated flux. Using these parameters, as in Ref.~\onlinecite{GueroutPRB12}, we study the heat transfer $\varphi(h)$ as a function of the grating height $h$, and discuss the amplification factor $\varphi(h)/\varphi(h=0)$ with respect to the case $h=0$, i.e. to the case of a planar slab. The results are shown in the main part of Fig.~\ref{Figh} for a very wide range of values of $h$, going from $h=0$ to 1\,mm, and compared with the ones (red points) calculated in Ref.~\onlinecite{GueroutPRB12} from $h=0$ to 6$\,\mu$m. While for $h=100\,$nm the amplification factor is close to 1, increasing $h$ produces a huge amplification of the heat flux. In particular, the ratio $\varphi(h)/\varphi(h=0)$ grows monotonically with $h$, and reaches a horizonal asymptotote around $h=500\,\mu$m, with a value slightly above 34. In the inset of Fig.~\ref{Figh}, we show the same curve in linear $h$ scale from 0 to $6\,\mu$m. As manifest from the plot, in spite of a qualitative agreement, the two curves are quantitatively different, with a disagreement going up to around 50\% for $h=5\,\mu$m. For this value of the grating height we have calculated the flux using FMM with $N=51$ and obtained a result in agreement within around 1\% with the one coming from the ASR with $N=8$. We conclude that the truncation order $N=51$ is large enough to assure a good convergence for the integrated flux using FMM. Thus, the difference between our results and the ones of Ref.~\onlinecite{GueroutPRB12} does not seem to originate from an insufficient truncation order.

It is now instructive to investigate the evolution of the spectral properties of the heat flux as a function of the grating depth $h$. This behavior is illustrated in Fig.~\ref{FigGrid}. Panel (a) shows the spectral flux $\varphi(\omega)$ for $h=0$, i.e. in a slab-slab configuration. It is well known that in presence of surface modes, while approaching the near-field regime the heat flux becomes more and more monochromatic at the resonance frequency of the mode~\cite{JoulainSurfSciRep05}. Even if the distance $d=1\,\mu$m does not fully lie in this distance regime, a signature of the existence of such surface modes is already present. Nevertheless, this is not the case for two gold slabs, simply because the plasma frequency for gold is located well outside the window of frequencies contributing to the flux, defined by the Planck function $n(\omega,T)$. As a result, the only non-monotonic behavior for the flux is observed in the low-frequency region $\omega\in[10^{10},10^{13}]\,$rad/s, shown in the inset of Fig.~\ref{FigGrid}(a). The spectral flux tends to increase at small frequencies as a result of the divergence of $\varepsilon(\omega)$ for $\omega\to0$, this growth being at some point compensated by the fact that each mode of the field carries an energy $\hbar\omega$, tending to 0 for $\omega\to0$.

Let us now focus on Fig.~\ref{FigGrid}(b), where the grating heights $h=1\,\mu$m and $1.5\,\mu$m are taken into account. We first remark that the non-monotonic behavior at low frequencies is still present, but with a decreased maximum value. This feature remains basically unchanged for all the higher values of $h$ shown in Fig.~\ref{FigGrid}. The main feature of Fig.~\ref{FigGrid} is anyway the appearance of two peaks in the spectral flux, sign of the existence of two surface modes, whose frequency depends on the grating height, coherently with the fact that these are indeed spoof plamons. The participation of these new modes produces an overall flux amplification equal to 2 and 2.7, respectively, for $h=1\,\mu$m and $1.5\,\mu$m. An analysis of panel (c) of the same figure shows that increasing $h$ produces both a decrease of the frequency of spoof-plasmon modes and an amplification of the peak height. As shown in Fig.~\ref{FigGrid}(c)-(d) the scenario becomes even more interesting for higher values of $h$. In this case more and more resonances enter the window of frequencies relevant for radiative heat transfer. This feature has been already discussed in Ref.~\onlinecite{GueroutPRB12} by analyzing the heat-flux transmission coefficient for a given value of the wavevector $\mathbf{k}$. We basically observe two resonances for $h=4\,\mu$m and $7\,\mu$m, while for $h=10\,\mu$m and $20\,\mu$m not only we start observing a comb of resonance frequencies, but we clearly see that the fact that they approach each other produces a constructive interference between them, increasing the overall value of $\varphi(\omega)$ in the frequency range considered. This is even more evident in Fig.~\ref{FigGrid}(f), where for $h=100\,\mu$m the frequency-comb behavior is manifest, as well as how this eventually produces a smooth asymptotic $\varphi(\omega)$ (reached for $h=1\,$mm) which does not show any abrupt change with respect to $\omega$. This is the profile giving the asymptotic flux amplification close to 34 discussed above and shown in Fig.~\ref{Figh}.

In order to compare further our results with previous works, we focus on a configuration studied in Ref.~\onlinecite{DaiPhD}, namely the heat transfer between two identical gratings having height $h=4.7\,\mu$m, period $D=500\,$nm and filling factor $f=0.6$, placed at distance $d=1\,\mu$m. The two chosen temperatures are $T_1=300\,$K and $T_2=301\,$K. In Ref.~\onlinecite{DaiPhD}, the authors show the spectrum associated with this heat transfer (calculated using the standard FMM) from which we have extracted some points, shown in red in Fig.~\ref{FigSpectrum}. We have applied our numerical scheme to this scenario and obtained the black curve shown in the same figure. It is manifest that, while sharing the same qualitative behavior, the two curves considerably differ quantitatively, in particular in the region between the two main peaks. In order to try to explain this disagreement we have noticed that in one of their recent papers on this topic (namely, Ref.~\onlinecite{DaiPRB2016b}), the authors clearly state that the fact that the two structures coincide implies that the two reflection operators $\mathcal{R}^{(1)+}$ and $\mathcal{R}^{(2)-}$ are equal. As discussed above, this is actually not correct because of the fact that the unit polarization vector in TM polarization depends on the propagation direction $\phi$. Nevertheless, we have performed a new calculation for the same structure, using the ASR with $N=8$, but omitting the sign change we need to deduce $\mathcal{R}^{(1)+}$ from $\mathcal{R}_u^{-}$. The result so obtained is given by the blue dashed curve in Fig.~\ref{FigSpectrum} and clearly shows an impressively increased agreement with the red points. Thus, we interpret the difference between the two results as due to this sign change, missing in Ref.~ \onlinecite{DaiPhD} and related works. Moreover, it must be stressed that the results presented in Ref.~\onlinecite{DaiPhD} also show an oscillatory behavior in frequency, in particular in the region between the two main peaks, which is completely absent in the results obtained by exploiting the ASR, both with and without the sign change. In our opinion, this oscillations are a result of the well-known instabilities of the FMM when dealing with metals.

With the aim of further trying to explain the discrepancy with Ref.~ \onlinecite{GueroutPRB12} shown in Fig.~\ref{Figh}, we have recalculated using the ASR one of the points of Fig.~\ref{Figh}, the one having $h=4.5\,\mu$m, without the mentioned sign change. The result is the blue dot in Fig.~\ref{Figh} which, even if approaching the result of Ref.~\onlinecite{GueroutPRB12}, still does not show a good agreement.

\section{Conclusions}\label{SecConcl}

We have addressed the caclulation of the radiative heat flux between two gold gratings. To this aim we have made use of the ASR, a modified version of the FMM introduced to deal with the high dielectric contrast typical of metals. We have shown that this technique produces a striking increase of the convergence rate, allowing us to obtain the heat flux at a distance of 1\,$\mu$m with a truncation order as low as $N=8$. This implies a remarkable gain in computational time: to give an idea, for the grating parameters relative to Fig.~\ref{FigFMMASR}, the calculation of the spectral flux at frequency $\omega=10^{14}\,$rad/s takes approximately 1 minute using the ASR with $N=8$ on a parallel 3.4\,GHz 16-core machine, while 1.5 hours are needed using the FMM with $N=51$. Since the required truncation order increases when decreasing the distance, the ASR would allow to explore smaller separations, which would be prohibitive using the FMM.

By using this improved numerical method, we have made a detailed study of the heat-flux amplification as a function of the grating depth. By studying both the flux ratio and the spectral properties, we have proved for our structure an amplification factor going up to 34, explained in terms of the appearance of a comb of spoof-plasmon resonances. We have then compared our results to some previous works, and highlighted a quantitative disagreement both in the integrated flux and in its spectral distribution. In some cases, we have argued that this disagreement is due to the incorrect assumption that the reflection matrices of the two gratings coincide.

Our results show that the physics behind metallic gratings makes them ideal candidates for the manipuation of both the overall heat flux and its frequency components, both possibilities being very relevant for several applications such as thermophotovoltaic energy conversion. On a more technical side, our discussion proves that the FMM can provide slowly-converging and unstable results in the presence of metals, while the ASR represents a much more reliable approach even for high dielectric contrasts.

\begin{acknowledgments}
The authors acknowledge J. Dai for useful discussions.
\end{acknowledgments}

\end{document}